# Architected Lattices with Adaptive Energy Absorption

*Yifan Wang[1,+], Brian Ramirez[1,2,+], Kalind Carpenter[3], Christina Naify[3,4],
Douglas C. Hofmann[3] & Chiara Daraio[1]*

[1]*Division of Engineering and Applied Science, California Institute of Technology, Pasadena, California 91125, USA.*
[2]*Department of Mechanical Engineering, California State Polytechnic University, Pomona, California, 91768, USA*
[3]*Jet Propulsion Laboratory/California Institute of Technology, Pasadena, California 91109, USA*
[4]*U.S. Naval Research Laboratory, Washington, DC 20375, USA*
[+]*Y. W. and B. R. contributed equally to this work*

**ABSTRACT:** **Energy absorbing materials, like foams used in protective equipment, are able to undergo large deformations under low stresses, reducing the incoming stress wave below an injury or damage threshold. They are typically effective in absorbing energy through plastic deformation or fragmentation. However, existing solutions are passive, only effective against specific threats and they are usually damaged after use. Here, we overcome these limitations designing energy absorbing materials that use architected lattices filled with granular particles. We use architected lattices to take advantage of controlled bending and buckling of members to enhance energy absorption. We actively control the negative pressure level within the lattices, to tune the jamming phase transition of the granular particles, inducing controllable energy absorption and recoverable deformations. Our system shows tunable stiffness and yield strength by over an order of magnitude, and reduces the transmitted impact stress at different levels by up to 40% compared to the passive lattice. The demonstrated adaptive energy absorbing system sees wide potential applications from personal protective equipment, vehicle safety systems to aerospace structures.**

Architected lattices are materials whose properties arise from the selection of both their constitutive materials and the geometry of their micro- and meso-structure [1-5]. Architected materials have been proposed as new energy absorbing solutions with recoverable deformation, for example, taking advantage of mechanical instabilities in their underlying structure [6-8]. Although reusable, these solutions are intrinsically passive, with properties fixed once fabricated, and effective in mitigating impact loads under predefined velocities or energies. Most practical applications, however, require adaptive structures whose mechanical properties can be tuned to absorb or dissipate varying amounts of energy, in response to different impact conditions.

Solutions to tune the mechanical properties of materials and structures [9, 10] include the use of hydrogels that respond to temperature, pH, light and water content [11,12]; shape memory alloys and polymers (SMAs and SMPs) [13,14]; liquid crystal elastomers (LCEs) that respond to temperature and light [15,16]; and magnetorheological (MR) and electroactive polymers (EAPs) [17-19]. However, these materials either are mechanically too soft for engineering applications (hydrogels), require large temperature changes (LCEs), need re-programming at high temperatures (SMAs and SMPs), or require strong electromagnetic fields (MR materials, EAPs), which are not easily accessible in most practical scenarios.

Granular systems are known to exhibit tunable mechanical properties during jamming, when the packing fraction of the particles is increased [20-24]. Jamming is a phase transition that does not rely on temperature changes, like in ordinary materials, but it is instead controlled by local geometric constraints. When a granular system jams, it undergoes a sharp transition from a soft to a rigid state, with large increases in stiffness and yield stress (usually more than an order of magnitude) [25]. The jamming transition in granular materials has been employed in engineering applications such as soft robotics [26] and granular architectures [27]. In this work, we design architected lattices with hollow members, which we then fill with granular particles (**Figure 1**a). We utilize the jamming phase transition of the filling particles, to create architected lattices with adaptive mechanical behavior. The effective constitutive response of the lattices is controlled applying different, negative gauge pressures within the members of the lattice, which jam the particles (**Figure 1**a). This design solution is particularly suitable for energy absorbing materials, because it leverages both the energy mitigation mechanisms of structured lattices (i.e., the buckling and bending of the members) and the frictional energy dissipation mechanism of granular systems.

In impact attenuation systems, the goal is to absorb or dissipate the incoming impact energy while maintaining the load outcome below an acceptable threshold level. Typical foams and lattices have a characteristic stress-strain response to compressive loads (**Figure 1**b), which includes elastic, yielding and densification regimes [1]. The amount of energy absorbed or dissipated by a lattice can be evaluated as the area under the stress-strain curve. The best energy absorbing materials are the ones that absorb a large amoung of incoming energy while transmitting the lowest stress. Since most constitutive responses of energy absorbing materials are nonlinear [1], the ideal material response differs as a function of the impact energy. As an example, let's consider two conceptual scenarios: (i) At low impact energy, a stiff lattice with a high yield stress (**Figure 1**b left, red curve) will accommodate impact energy in its elastic region, but may reach a high transmitted stress. Softer foams (**Figure 1**b left, green or blue curves), will be able to dissipate the same amount of energy while undergoing larger deformations in their yielding regime, transmitting a lower stress. (ii) At higher impact energies (**Figure 1**b middle and right), however, softer foams may undergo too large deformations, reaching the densification regime, where the transmitted stress increases dramatically. For these scenarios, stiffer foams perform better. Hence, a particular foam is usually performing optimally only in a specific range of impact energies. In our architected lattices, the energy absorption and transmitted stress can be dynamically tuned to best perform based on the expected impact energy.

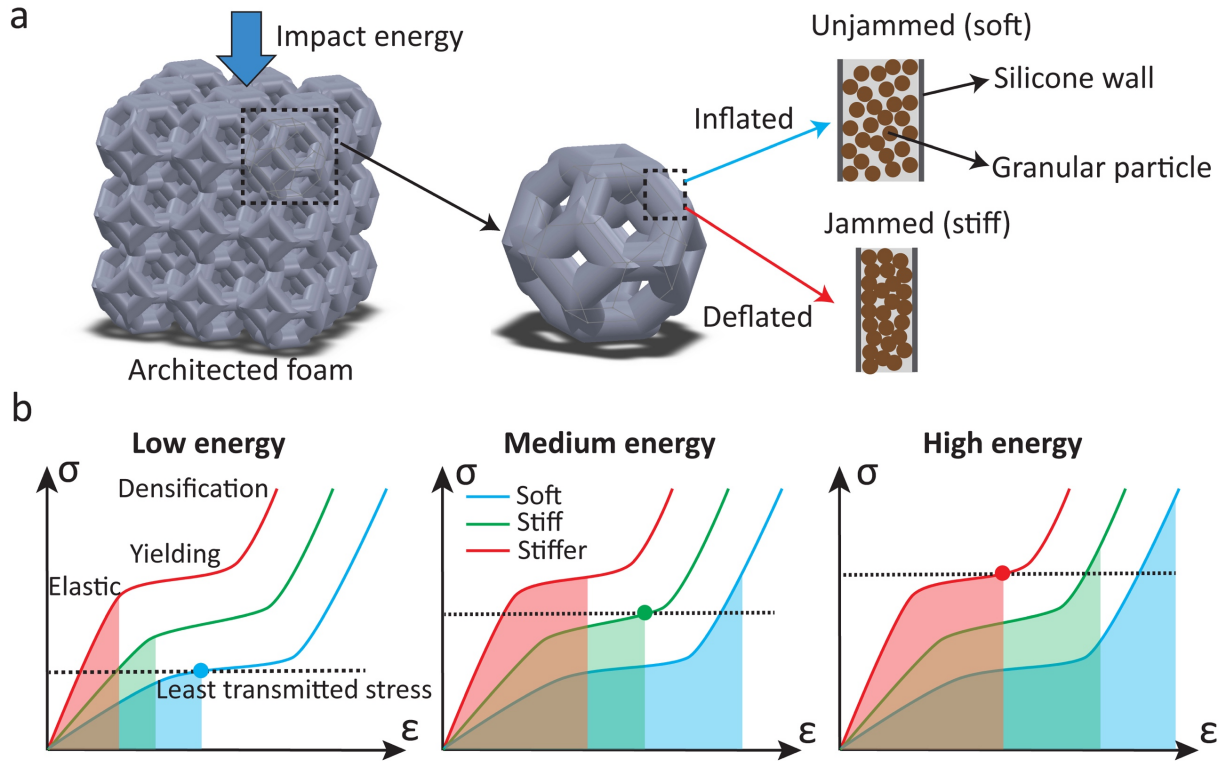

**Figure 1.** a) Schematic of an architected lattice (left) composed of a cuboctahedron (Kelvin) unit cell (middle). The architected cell consists of 3D printed hollow silicone struts with granular particles filled inside (right). The cell struts can change volume by controlling their internal pressure, leading to jamming phase transition in the granular materials inside. b) Schematic of different stress vs. strain curves for impact absorbing materials with different effective stiffnesses, subjected to low, medium and high impact energies (the colored areas show equal energy absorption by the three foams in the different impact scenarios). The horizontal dashed line shows the lowest transmitted stress. The schematic shows that the mechanical response of absorbing materials need to be tuned according to impact energy, to minimize impact stress transmission.

To characterize the quasi-static response of the constitutive elements of our lattices, individual hollow strut members (**Figure 2**a) were fabricated using flexible silicone material and filled with granular particles (ground coffee, see *Experimental Section*). A negative internal gauge pressure is applied to the struts, using a pump connected to the inner volume through a thin tube. As the negative pressure is varied, the volume of the strut decreases by ~5% at 40 *kPa* and by ~15% at 93 *kPa*. This volume contraction leads to an increase in the granular packing fraction, eventually causing the jamming phase transition. In order to obtain quantitative information on the mechanical properties evolution between the unjammed and jammed state, the strut elements were tested under varying internal negative pressures (**Figure 2**a). Due to the complex and anisotropic mechanical behavior of granular materials, we adopted a simplified composite strut model to capture the strut's elastic response. In the model, we assumed that the axial and bending modes of deformation of the strut can be decoupled, similar to previous studies on MR composite lattices [18]. The effective Young's modulus ($E$) of the strut is characterized under quasi-static uniaxial compression tests as a function of internal negative pressure ($P$). The effective bending modulus ($E_{bend}$) was measured with 3-point bending tests and was related to the shear modulus ($G$) by $E_{bend} = \frac{9KG}{3K+G} \approx 3G$, assuming the bulk modulus ($K$) is much larger than $G$ [18]. As the internal negative pressure ($P$) increases from 0 to 93 kPa, the Young's modulus $E(P)$ increases from 0.11 MPa to 1.85 MPa and the shear modulus $G(P)$ increases from 0.05 MPa to 0.83 MPa (**Figure 2**a). These large changes in moduli, by over an order of magnitude, surpass most other stiffness-changing materials [10].

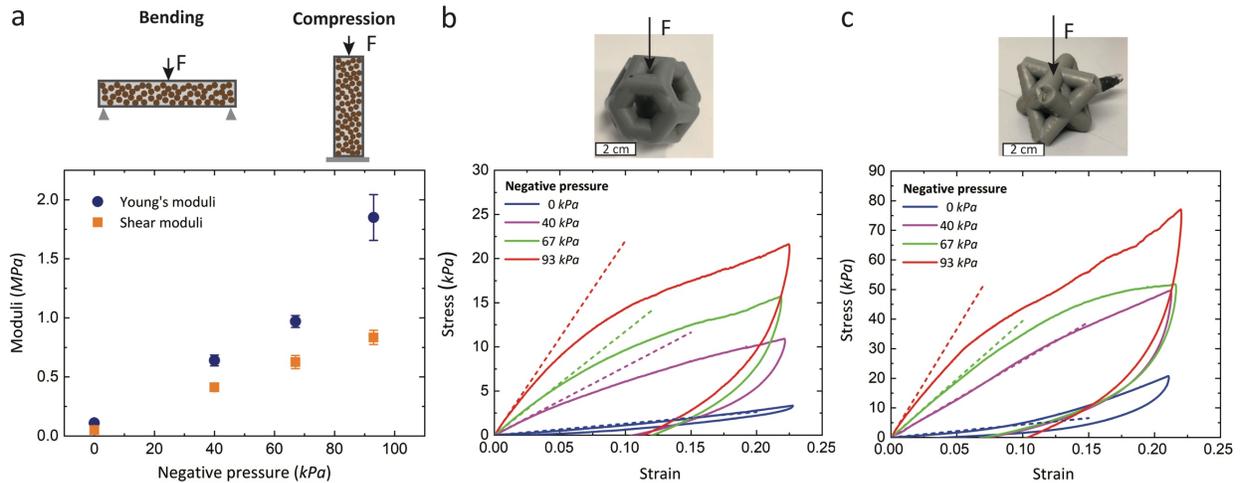

**Figure 2.** a) Variation of the effective Young's modulus ($E$) and shear modulus ($G$) of individual struts tested under different internal negative pressure conditions. (top) Schematic of the 3-point bending and uniaxial compression tests performed on a single member. b) Compression stress vs. strain data for the bending-dominated (Kelvin) cell at different negative pressures. c) Compression data for the stretching-dominated (Octet) cell at different negative pressures. The dashed lines in (b) and (c) show the lattices' effective loading stiffness predicted by the finite element simulations, using data from (a).

We then fabricated structured lattices filled with granular particles. We selected two representative geometries for the lattice architecture: the bending-dominated (Kelvin, **Figure 2**b) cell and the stretching-dominated (Octet, **Figure 2**c) cell, to represent two extreme cases in the stiffness-density scaling diagram [1, 3]. The Kelvin and Octet single unit cells were then tested under quasi-static uniaxial compression, varying their internal negative pressure (**Figures 2**b, c). It is important to note that the architected unit cells yield quickly after the initial elastic region, which is ideal for energy absorption, as in this regime energy is dissipated into heat by the frictional flow of the granular particles. The fact that the stiffness and yield stress can be varied controlling the internal pressure is useful to tune the material's energy absorption to specific impact threats. The quasistatic results on the two different unit cells also show that the Kelvin cell has much lower stiffness and yield strength compared to the Octet cell, at similar relative densities (see *Experimental Section*), due to its bending-dominated deformation mode and buckling behavior upon yielding.

In order to inform the mechanical response of the architected cells, we develop a simplified finite element model. The model assumes the architected cells are constructed from struts with homogenized, linear elastic responses. These struts are then modelled with pressure-dependent elastic parameters *E(P)* and *G(P)*, obtained from experiments (**Figure 2**a and *Experimental Section*). The numerically calculated elastic loading regimes (dashed lines in **Figures 2**b, c) match well the experimental data. With this model, the quasi-static elastic response of different cell geometries can be predicted and used to guide future designs.

Because of its buckling behavior, we select the Kelvin cell to perform dynamic impact tests. We characterize the cell under different impact loading conditions, using a drop-weight tester (**Figure 3**a, and *Experimental Section*). We reconstruct stress-strain relations, $\sigma(\varepsilon)$, for unit cells subjected to different negative pressures (**Figures 3**b, c, and *Experimental Section*). At low impact velocity (1 *m/s*, **Figure 3**b), the unjammed cell (at 0 *kPa* negative pressure) provides the most efficient energy absorption, reaching the lowest transmitted peak stress. However, when impacted at higher velocities (3 *m/s*, **Figure 3**c), the same cell undergoes larger deformations, causing self-contact between struts. This causes the peak stress to drastically increase. Effective protection at higher impact velocities calls for an increase in stiffness and yield strength. At 3 *m/s*, the cell subjected to a 93 *kPa* negative pressure reaches the lowest transmitted peak stress. As stiffness and yield strength of the architected lattices are tuned, they achieve greater stress attenuation.

For a given lattice geometry, we compare peak transmitted stress at different impact velocities (varied between 1 *m/s* to 3.7 *m/s*), for different internal pressures (**Figure 3**d). We create a design map (shaded regions in **Figure 3**d), to identify the negative pressure required to minimize the transmitted impact stress and to obtain improved energy absorption at each impact velocity. The map clearly shows that higher impact velocities require higher negative pressures, to increase the overall effective stiffness of the unit cell. We assess the energy absorption efficiency of the architected materials by plotting their cushioning efficiency ($C_E$) [8], defined as the ratio between the energy input (*W*) and the load stress produced (σ),
$$C_E = W/\sigma \ . \tag{1}$$
For cushioning materials, $C_E$ is usually expressed in geometry independent terms (i.e., strain energy density per unit stress) and its value varies between zero and one. Efficiency generally increases up to the onset of densification, then declines at higher strains. Among passive cushioning foams, protective materials will generally be selected for higher $C_E$, at a specific load and/or impact energy. In order to create cushioning systems that are effective in broader loading ranges, materials with different properties are usually layered together. However, such layered systems generally reduce the overall protective performance, because the different layers reach optimal $C_E$ at different load levels. **Figure 3**E shows the cushioning efficiency curves for our lattices, at different negative pressures. Each curve, reaches peak cushioning efficiency at different load stress, a response similar to that of traditional foams with different densities The measured peak cushioning efficiency varies from 0.2 to 0.25, which falls within the range of 0.2 to 0.4 reported for 3D architected structures and polymeric foams [28, 29]." In the passive structures and foams, the mechanical properties and cushioning efficiency are fixed and one must choose the appropriate density for the given application. However, in our lattice material, the same lattice can be used for different load ranges, by tuning the negative pressure to the highest cushioning efficiency. Essentially, the cushioning efficiency of our material can be tune to any value within the envelope of the individual curves, obtained for different negative pressures

    

(black dashed line in **Figure 3**e). Therefore, our material can obtain high efficiency across a wider range of impact energies, compared to a passive material that is limited to a single peak of efficiency, at a particular impact load or energy.

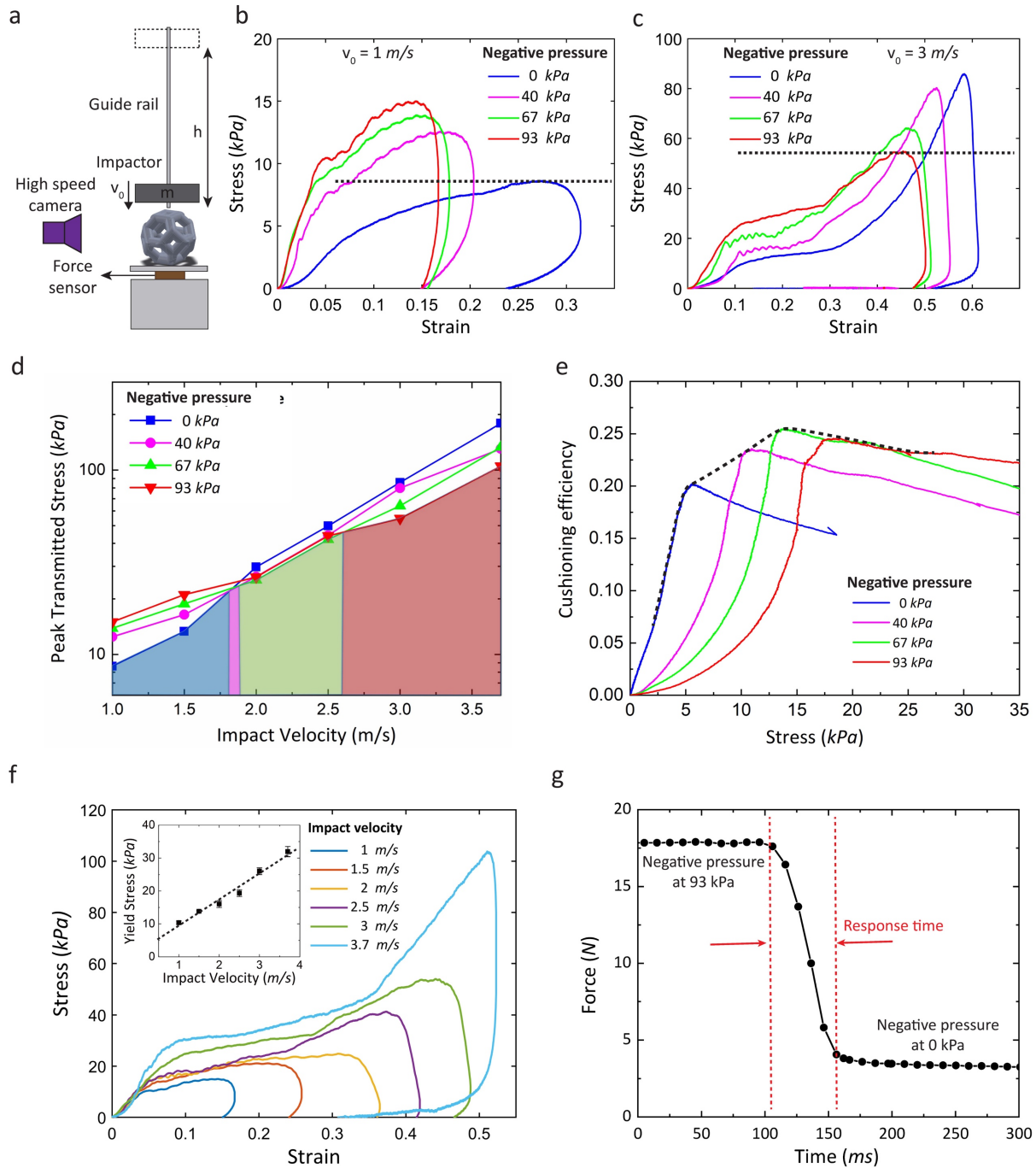

**Figure 3.** a) Schematic of the experimental setup for drop-weight impact tests on a Kelvin cell. b) Stress-strain curves at 4 different negative pressures with impact velocity $v_0 = 1\ m/s$. The cell with zero negative pressure (0 kPa) shows lowest peak stress transmission (black dashed line). c) Stress-strain curves at impact velocity $v_0 = 3\ m/s$. The cell with a 93 kPa negative pressure shows lowest peak stress

transmission (black dashed line). d) Peak transmitted stresses for different impact velocities and negative pressures. The colored areas indicate the negative pressures that minimize stress transmission, at varying impact velocities. e) Cushioning efficiency curves at different negative pressures. The envelope of these curves (black dashed curve) is the reachable cushioning efficiency with active pressure tuning. f) Stress-strain curves for the Kelvin cell, confined by 93 *kPa* negative pressure and impacted at different velocities. The inset shows the yield stress under these varying impact velocities. g) The force-time response of the Kelvin cell under compression, when the negative pressure of 93 *kPa* is suddenly removed. The dashed red lines shows a response time on the order of ~50 *ms*.

The Kelvin cell filled with granular particles shows a rate-dependent behavior. At constant negative pressure, the force-displacement curves demonstrate an increase in yield stress with increasing loading rate (**Figure 3**f), a behavior similar to non-Newtonian fluids [30]. This rate sensitivity can be attributed to an increase in local densification and jamming of the granular particles. At higher impact velocities, the rate of deformation is so high that the granular particles inside the struts are not fast enough to flow, a phenomenon already observed in bulk granular systems under impact [31]. At lower velocities, the granular particles are given enough time to flow, reducing the overall yield stress.

In order to characterize the response time of our architected lattice, we measured the load-time response of the cell while changing its internal pressure. For this, we pre-compressed a Kelvin cell in an Instron materials testing machine to a 12% strain and held displacement constant. The internal negative pressure was then removed, going from 93 *kPa* to 0 *kPa.* The dynamic load change was recorded at a rate of 100 points/second (see *Experimental Section*). The measured force shows a decay step, with transition period of ~50 *ms* (**Figure 3**g). This response time is much shorter than that observed in responsive metamaterials filled with MR fluids (~1s) [18] and it is similar to light activated liquid crystal elastomers [32]. It is important to note that the response time is on the same scale as the impact duration (30 – 50 *ms*), demonstrating the ability of the architected cell to adapt its stiffness within the impact period.

Most lightweight energy absorbing materials are efficient in absorbing impact energy through plastic deformation or crushing, but are unable to recover their shape and properties after an impact. Our architected lattices are able to completely recover their original state after impact, if the internal negative pressure is removed (**Figure 4**a). Different unit cells were subjected to a series of 50 impacts with a 5.0 seconds intervals. Negative pressure was removed and re-applied after each impact. The architected unit cells showed no degradation of their mechanical properties. This ability to recover after impacts is essential for applications in scenarios where multiple impacts are expected and reusability is required.

Finally, we demonstrate the ability to scale up the architected cells into larger lattices composed of $2 \times 2 \times 2$ unit cell arrangements, filled with granular particles. These lattices can also be tuned varying the internal pressure within their members. As an example, we show that the $2 \times 2 \times 2$ Kelvin lattice varies its stiffness from a rigid and load bearing state (**Figure 4**b) to a soft state (**Figure 4**c). In both cases, the lattice support a 3.5 *kg* external load, but its deformation is significantly different – the strain varies from 2.6% in the rigid state, to 25% in its softer counterpart. Quasi-static compression tests were also performed on these lattices, to quantitatively characterize the change in mechanical properties (see *Supplementary material*). The measured stress-strain relations at different internal pressures qualitatively match with those observed in the single cell measurements, with small differences resulting from a change in the boundary conditions.

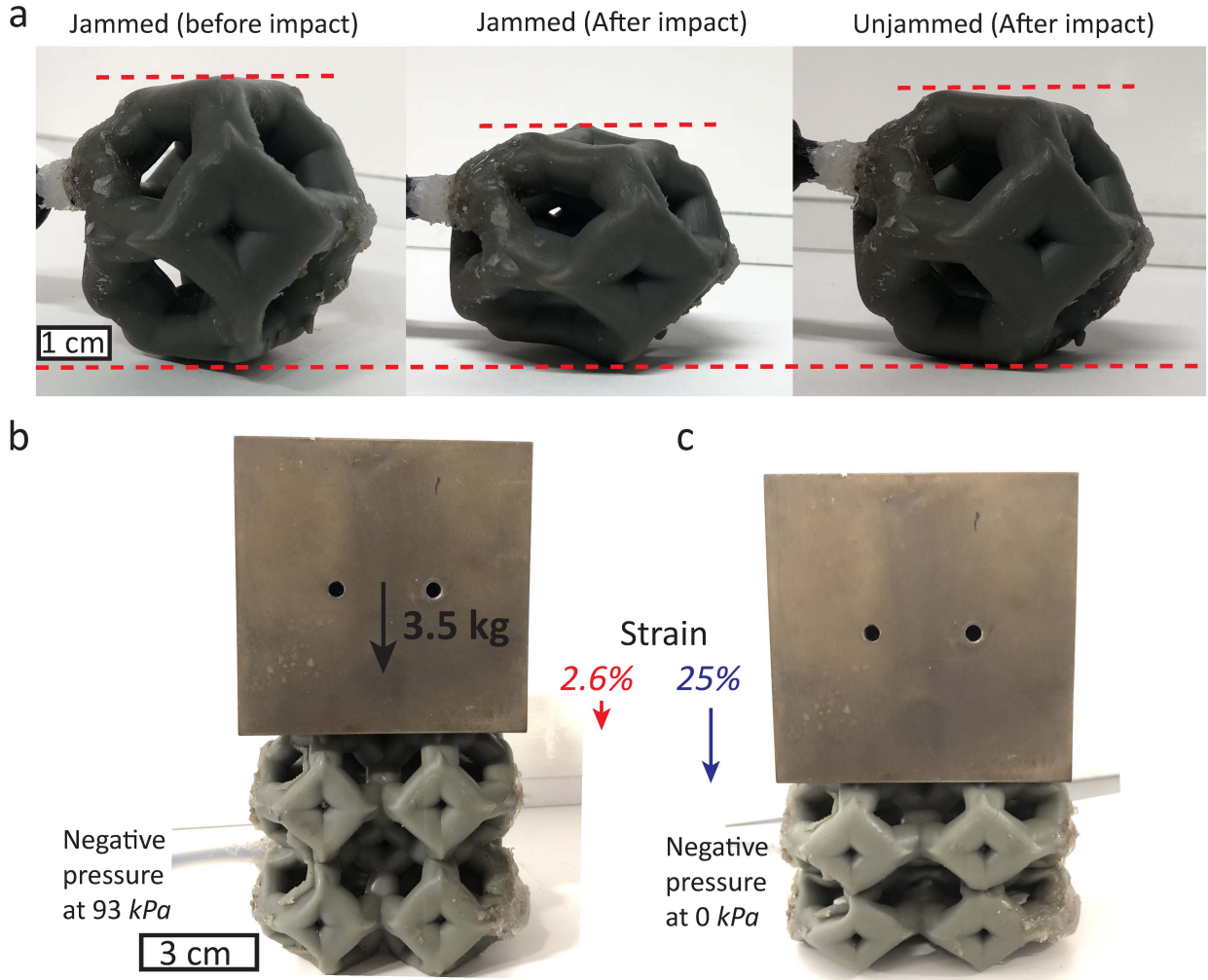

**Figure 4.** a) Images of the Kelvin cell when jammed before impact (left), jammed after impact (middle), and unjammed after impact (right). The fact that the cell returns almost to original height demonstrates the recoverability after impact. b) An architected 2 × 2 × 2 Kelvin lattice composed of Kelvin unit cells in the stiffer state, when a negative 93 *kPa* pressure is applied, holding a steel block of 3.5 *kg*. The measured uniaxial strain is 2.6%. c) The same architected lattice in soft state at zero negative pressure (0 *kPa*), holding a 3.5 *kg* steel block. The measured uniaxial strain is 25%, about 10 times higher than in (b).

In this work, we demonstrated the first adaptive, architected material, with tunable energy absorption. The effective properties can be tuned to minimize stress transmission, while retaining peak cushioning efficiency over a much broader span of loading conditions, compared to conventional passive materials. The proposed tuning mechanism relies on jamming transitions in the granular filling. As such, it can function with different granular materials, which can be

Copyright California Institute of Technology 2019    8

selected based on the needs of applications. Recent advances in additive manufacturing make it possible extend our concept to different scales, suggesting its use for different applications, ranging from structural cushoning to wearable protective gear.

**Experimental Section**

*Materials:* Hollow strut members and lattice structures (**Figures 2**a-c) were fabricated using a flexible silicone material (Sil-40) in a Carbon3D© commercial 3D printer. The manufacturing process and materials were selected based on the ability to create high-resolution, hollow unit cells and lattices capable of large elastic deformation, while simultaneously being able to retain granular materials inside and hold a tight pressure seal. Both cell sizes are $40 \times 40 \times 40$ mm and the hollow struts are 10 mm in outer diameter and 8.4 mm in inner diameter. After the hollow struts and lattices are printed, granular particles (ground coffee, by Verena Street) are filled into the hollow cells by a funnel through a pre-printed vent hole. The density of the granular particles in the lattices without negative pressure is measured to be $510 \text{ kg}/m^3$. Ground coffee is chosen as the granular material because of its good performance in jamming hardness tests and relatively low density [26], while other types of granular particles can also be used and are not expected to change the main conclusions of this paper [33]. The resulting composite foam has an equivalent density of $320 \text{ kg}/m^3$ for the Kelvin cell and $312 \text{ kg}/m^3$ for the Octet cell, comparable to commercially available foam products [8]. After filled, the vent hole is connected to a vacuum pump and then sealed with silicone glue in order to form an air-tight pressure controlled system.

*Quasi-static compression tests:* The fabricated struts and lattices are characterized under quasi-static uniaxial compression and 3-point bending tests, respectively, using an Instron E3000 materials tester, at a loading rate of 0.5 mm/s. 3 separate tests are repeated for each sample at the same internal pressure and the error bars in **Figure 2** represent the standard deviation. To obtain the response time, a cell was pre-compressed to a 12% strain and the displacement was maintained while the negative gauge pressure was suddenly removed from 93 *kPa* to 0 *kPa*. In order to perform a sudden removal of the negative pressure and separate the influence from vacuum pump speed, we connected the architected cell to a 3.8 Liter vacuum chamber pre-pumped below 0.1 *kPa*, whose volume is ~$10^2$ larger than the cell. The Instron testing system records data at a rate of 100 points/second, which is fast enough to capture the stiffness change from the architected cell (**Figure 3**g).

*Impact tests:* Specimens of $40 \times 40 \times 40$ mm$^3$ were placed on a flat force plate and impacted using a 45 mm-diameter copper cylinder of mass m $= 312$ g. The impact velocity $v_0$ was varied by changing the cylinder's initial drop height *h*. A piezoelectric force sensor was placed underneath force plate to measure the transmitted force *F* as a function of time *t*. The force-time data *F(t)* were then integrated to obtain the velocity *v(t)* and displacement of the impactor *x(t)*:

$$v(t) = v_0 - \int_0^t \frac{F(t)}{m} dt, \quad x(t) = \int_0^t v(t) dt \qquad (2)$$

 

The impact process was also recorded using a high speed camera (Phantom Vision Research) at a rate of 3000 frames/second, to track the position of the impactor during impact and compared with the integration results from above.

*Finite Element Simulations:* The pressure dependent Young's moduli $E(P)$ and shear moduli $G(P)$ obtained from the strut element tests were used as parameter inputs into the commercial finite element package ABAQUS/Standard for a 3D stress analysis. The models were meshed using 10-node quadratic tetrahedral elements (type C3D10H), with mesh sensitivity analysis performed to ensure accuracy. Uniaxial compression tests were performed in the finite element model (**Figures 2**b, c dashed lines) and compared with experimental measurements (**Figures 2**b, c solid lines).

**Acknowledgements**

Y. W and C. D. acknowledge the support from the National Science Foundation under EFRI Grant No. 1741565. This research was carried out at the California Institute of Technology and the Jet Propulsion Laboratory under a contract with the National Aeronautics and Space Administration, and funded through the President's and Director's Fund Program. Y. W., B. R. and C. D. designed the experiment. Y. W. and B. R. performed the experiment and analyzed the data. Y. W., B. R. and C. D. wrote the manuscript. All authors interpreted the results and reviewed the manuscript.

# Supplementary Material for

*Yifan Wang[1,+], Brian Ramirez[1,2,+], Kalind Carpenter[3], Christina Naify[3,4],*
*Douglas C. Hofmann[3] & Chiara Daraio[1]*

[1]*Division of Engineering and Applied Science, California Institute of Technology, Pasadena, California 91125, USA.*
[2]*Department of Mechanical Engineering, California State Polytechnic University, Pomona, California, 91768, USA*
[3]*Jet Propulsion Laboratory/California Institute of Technology, Pasadena, California 91109, USA*
[4]*U.S. Naval Research Laboratory, Washington, DC 20375, USA*
[+]*Y. W. and B. R. contributed equally to this work*

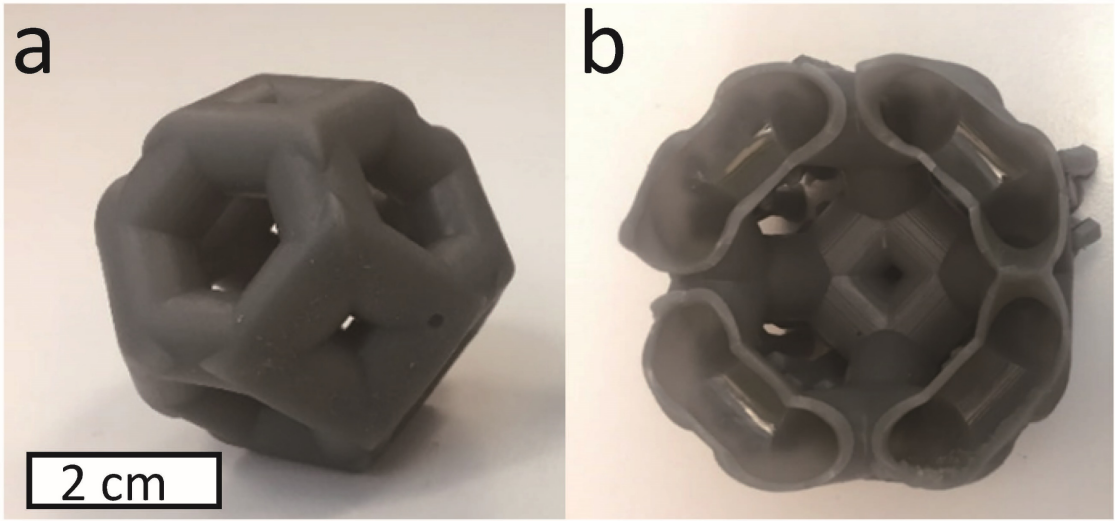

**Figure S1.** The fabricated unit lattice a) and a cut view of the members (hollow opening) b) to visualize the wall thickness and absence of support. The hollow lattice structures (**Figure 2**a-c) were fabricated using flexible silicone material (Sil-40) in a Carbon3D© commercial 3D printer. Both cell sizes are 40 × 40 × 40 mm and the hollow struts are 10mm in outer diameter and 8.4 mm in inner diameter.

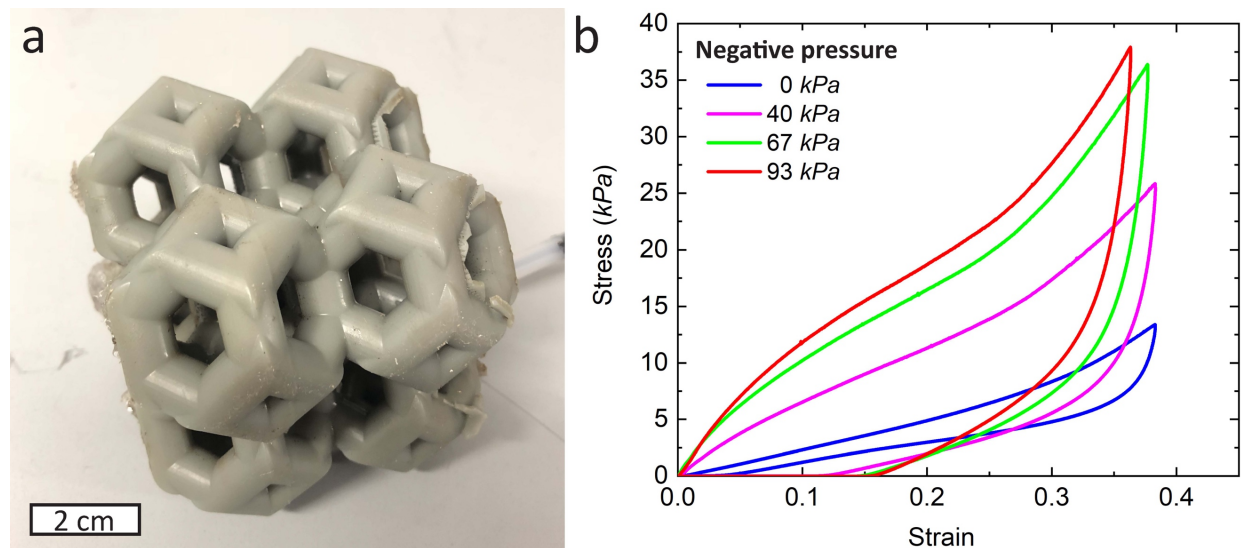

**Figure S2.** a) The fabricated 2X2X2 Kelvin lattice filled with granular particles. b) The quasi-static force-displacement curve for the lattice in (a) under different confinement pressures. The lattice shows similar stiffening behavior with the unit cell but with higher force levels.